\documentclass[aps,prb,twocolumn,groupedaddress,showpacs,floatfix]{revtex4}

\usepackage{amsmath}
\usepackage{graphicx}

\begin{document}

\title{ Microscopic mechanisms of dephasing due to electron-electron 
interactions }

\author{R. \v{Z}itko}
\author{J. Bon\v{c}a}

\affiliation{FMF, University of Ljubljana, 
and J. Stefan Institute, Ljubljana, Slovenia}

\date{\today}

\begin{abstract}
We develop a non-perturbative numerical method to study 
tunneling of a single electron through an Aharonov-Bohm ring 
where several strongly interacting electrons are bound. 
Inelastic processes and spin-flip scattering are taken into 
account. The method is applied to study microscopic mechanisms
of dephasing in a non-trivial model. We show 
that electron-electron interactions described by the Hubbard Hamiltonian
lead to strong dephasing: the transmission probability
at flux $\Phi=\pi$ is high even at small interaction strength. 
In addition to inelastic scattering, we identify two energy conserving mechanisms 
of dephasing: symmetry-changing and spin-flip scattering. 
The many-electron state on the ring determines which of these mechanisms
will be at play: transmitted current can occur either in elastic or 
inelastic channels, with or without changing the spin of the scattering electron.
\end{abstract}


\maketitle

\section{Introduction}

Advances in the semiconductor technology made it possible to study
quantum interference effects in mesoscopic systems where the wave
nature of electrons plays an essential role.  Particularly noteworthy
are the studies of the Aharonov-Bohm (AB) oscillations in mesoscopic
rings \cite{webb,timp,schuster}.  The analysis of results in terms of
the single-electron picture turned out to be inadequate to describe
the totality of phenomena.  Inelastic scattering of electrons is
believed to be the predominant mechanism responsible for the loss of the phase
coherence in such experiments and the suppression of the $h/e$
oscillations.  When an electron interacts with optical phonons, the
dephasing only occurs through inelastic processes \cite{bonca3}.  At
low temperatures the phonon degrees of freedom freeze out, therefore
other mechanisms for dephasing should be taken into
account. Measurements of the dephasing time saturation at low
temperatures \cite{mohanty1,mohanty2} show that zero-point
fluctuations of the electromagnetic environment
\cite{golubev1} could play a role in explaining
this anomalous behavior. It is nevertheless believed, that at low
temperatures the electron-electron interaction is the dominant
mechanism for dephasing \cite{mesoscopic,golubev2}. Further support for the
importance of electron correlations at very low temperatures comes from
recent measurements of anomalous temperature dependence of the
dephasing time in mesoscopic Kondo wires \cite{felicien} 
where non-Fermi-liquid behavior has been found below the Kondo temperature.

The AB geometries have been theoretically studied using self-consistent
mean-field approximations that break down for degenerate levels, which
physically happens at very low temperatures
\cite{phasechange,phasechange2,reflection}.  The mean-field
approximation does not describe transitions in which the symmetry of
the many-electron wavefunction of correlated bound electrons changes
and it is thus inadequate to study dephasing.  Keldysh type Green's
functions and numerical renormalisation group techniques
\cite{kondo,rg} or equation of motion method \cite{fanokondo} have
been applied to AB systems, where calculations were limited to
interacting quantum dot with two levels coupled to reservoirs.
Particular attention was devoted to the appearance of the Kondo
physics induced by changing the magnetic flux, however no spin-flip
induced dephasing has been investigated by these methods.  A study
of a Coulomb blockade regime was recently done by Xiong and Xiong
\cite{xiong} with a method, similar to the one proposed in the present
work. Their Hamiltonian, however, maps on a non-interacting model in
the limit when the coupling to the leads is zero.  Furthermore, they
have only investigated spinless fermions and neglected inelastic
scattering channels. Transmission of two interacting electrons was
recently studied on the basis of continuous two-particle Hamiltonian
where an enhancement of transmission with increasing interaction
strength was found \cite{aharony}.

To shed some new light on the problem of dephasing in electron
tunneling experiments, there is obviously a demand for a capable
method that would treat the problem of the scattering of an electron
through a finite region where electron-electron (e-e) interactions
would be exactly taken into account.  Such a method should be based on
the use of exact correlated many-electron wavefunctions. 

In this Paper we propose a method that treats the e-e interactions by
direct diagonalisation of the many-body Hamiltonian using iterative
(Lanczos) technique.  The method naturally takes into account
spin-flip processes, so it can predict the ratio of spin-flip over 
normal scattering processes. This makes the technique interesting for
calculating spin-polarized transport \cite{ab} in the field of
spintronics.

We apply the method to study single-electron transmission through an
Aharonov-Bohm (AB) ring with e-e interactions. As it is widely known,
an electron perfectly reflects from an AB ring when the flux $\Phi$
penetrating the ring is such that the phases gained by the electron
traveling through the lower or the upper arm of the AB ring cancel
each other ($\Phi=\pi$). Such reflection occurs for any energy of
the incident electron. This remains true even when there are
electrons bound on the AB ring, as long as the system remains
non-interacting. The main purpose of our investigation is in the
influence of the finite Coulomb repulsion on the transmission of the
electron in the case described above.  We choose the Hubbard model to
describe the AB ring. The Hubbard model is the simplest yet 
most important nontrivial prototype model for correlated
electrons in the solid state. As we will show, finite Coulomb
interaction in certain cases leads to finite transmission of the
incident electron, despite the fact that the total wavefunction
for the scattering electron and electrons bound on the AB
ring preserves full quantum coherence. We will therefore refer to the processes
that lead to finite transmission in the case when $\Phi=\pi$ 
as {\it dephasing} processes since they clearly lead to diminished 
AB oscillations observed in experiments.  
We show that dephasing can occur either by a) inelastic
processes where the tunneling electron excites bound electrons on the
ring or by b) elastic (in regard to energy) processes, where the
tunneling electron changes the symmetry or the spin of the degenerate
many-electron wavefunction.  No exchange of energy is required in the
latter case \cite{phase,model,ab,interferometry}: dephasing occurs
because the tunneling electron leaves a trace on its ``environment'',
which consists of bound electrons. 

\section{Method}

\newcommand{\bra}[1]{\langle {#1} |}
\newcommand{\ket}[1]{| {#1} \rangle}
\newcommand{\braket}[2]{\langle {#1} | {#2} \rangle}
\newcommand{\upa}{\uparrow}
\newcommand{\dna}{\downarrow}
\newcommand{\cc}[2]{c_{#1,#2}}
\newcommand{\cp}[2]{{c_{#1,#2}}^{\dag}}
\newcommand{\cpm}[2]{{a_{#1,#2}}^\dag}
\newcommand{\cpn}[2]{{b_{#1,#2}}^\dag}
\newcommand{\cm}[2]{{a_{#1,#2}}}
\newcommand{\cn}[2]{{b_{#1,#2}}}
\newcommand{\rng}{{\text{ring}}}
\newcommand{\lll}{{\text{L}}}
\newcommand{\rr}{{\text{R}}}
\newcommand{\fr}[2]{\frac{#1}{#2}}
\newcommand{\tk}{\tilde{k}}

The proposed method is based on the multichannel scattering technique
that was developed for studying the tunneling of a single electron in
the presence of scattering by phonons \cite{anda,bonca1}.  Since its
introduction, it has been successfully applied to a variety of
problems where a single electron is coupled to phonon modes
\cite{bonca2,bonca3,ness1,ness2,ness3,cp1,cp2,torres}. It was
even incorporated into Landauer theory
where the influence of electron-phonon scattering on the
non-equilibrium electron distribution has been investigated
\cite{Emberly}.  We now generalise this method to study many-electron
problems.

\subsection{Model Hamiltonian}

\begin{figure}[htbp]
\includegraphics[height=4.5cm]{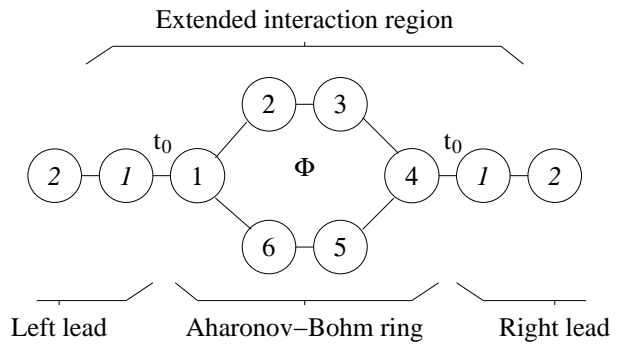}
\caption{Aharonov-Bohm ring. Magnetic flux penetrates the center of the
ring.\label{fig:ring}}
\end{figure}

While the method can be applied to more general situations 
and arbitrary geometries of the interaction region, 
we choose for simplicity a particular physical system which will 
also serve as a toy-model for the study of the e-e interaction induced
dephasing.  We thus consider an AB ring coupled to two ideal
one-dimensional leads, see Fig.~\ref{fig:ring}.  The ring is
described by a Hubbard-type Hamiltonian

\begin{equation}
\begin{split}
H_\rng = & \sum_{j,\sigma} \left( \epsilon\ \cp{j}{\sigma} \cc{j} {\sigma} 
- t e^{i \phi_j} \cp{j+1}{\sigma} \cc{j}{\sigma} + \textrm{h.c.} \right) \\
& + U \sum_j \cp{j}{\upa} \cc{j}{\upa} \cp{j}{\dna} \cc{j}{\dna}
\end{split}
\label{hring}
\end{equation}
The operator $\cp{j}{\sigma}$ creates an electron with spin $\sigma$
at site $j$ and we make a formal identification 
$\cp{7}{\sigma}=\cp{1}{\sigma}$.
The phases $\phi_j$ describe phase changes due to magnetic
flux penetrating the ring. We chose a gauge in which we ascribe
the total phase change due to magnetic field flux, $\Phi=2\pi \Phi_M/\Phi_0$,
(where $\Phi_M$ is the magnet field flux and $\Phi_0=h/e$ is the
flux quantum)
to a single element, {\sl e. g.} $\phi_1=\Phi$ and $\phi_j=0$, $j\neq 1$.

The leads are described by a tight-binding Hamiltonian
\begin{eqnarray}
{H_{\text{lead}}} = -t_{\text{lead}} \sum_{i,\sigma} \cpm{i+1}{\sigma} \cm{i}{\sigma} + \textrm{h.c.} \\
-t_{\text{lead}} \sum_{i,\sigma} \cpn{i+1}{\sigma} \cn{i}{\sigma} + \textrm{h.c.}
\end{eqnarray}
The operator $\cpm{i}{\sigma}$ creates an electron with spin $\sigma$ at site $i$
on the left lead, while the operator $\cpn{i}{\sigma}$ does the same on the right lead.
The ring is coupled to the electrodes with coupling constants $t_0$:
\begin{equation}
H_c = -t_0 \sum_{\sigma} \left( \cpm{1}{\sigma} \cc{1}{\sigma} + \textrm{h.c.} \right)
-t_0 \sum_{\sigma} \left( \cpn{1}{\sigma} \cc{4}{\sigma} + \textrm{h.c.} \right)
\end{equation}
The coupling constant $t_0$ need not be small: the method applies equally
well for strong coupling between the interaction region and the leads.

\subsection{Correlated many-electron states and exact diagonalisation}

The transmittivity can be meaningfully defined in many-electron
scattering problem only if one single electron leaves the scattering region.
For this reason, we restrict the energy of the incoming electron to be 
below the ionisation threshold.
Our approximation then consists of taking into account only
those many-electron states in which at most one (scattering)
electron is located outside the ring. 
Before the impact of the electron (which, for convenience,
will be chosen to have spin up), there are $n=n_\upa+n_\dna$ other
electrons bound on the AB ring. We truncate all many-body states,
where additional electrons hop from the interacting region to the
lead. When  physical parameters of the system, {\it e.g.}
$(\epsilon,t,U)$, are chosen in such a way that these $n$ electrons 
are bound in the interacting region, the approximation is equivalent
to neglecting the exponentially decaying tails of the $n$-electron
wavefunction in the leads. 

Before the scattering, the bound electrons are therefore in one of
the $n$-particle eigenstates of the Hamiltonian $H_\rng$,
Eq.~(\ref{hring}). We denote
these states by $\ket{\alpha_i^\upa}$ and their energies by $\epsilon_i^\upa$.
The superscript index $\upa$ denotes that the electron in the
lead has spin up.

When the incoming electron enters the ring, 
the system is in a superposition of the $n+1$-particle states which we 
denote by $\ket{\beta_i}$. These states are not necessarily 
eigenstates of $H_\rng$. 
After the scattering there is a single electron in
the leads, while the ring is in a superposition of the $n$-particle
eigenstates of $H_\rng$.
These states are the $\ket{\alpha_i^\upa}$
states and (if the spin of the scattering electron can be flipped)
the $n$-electron
eigenstates with $n_\upa+1$ spin-up electrons   and
$n_\dna-1$  spin-down electrons. These spin-flipped states are labeled 
by $\ket{\alpha_i^\dna}$ and their energies by $\epsilon_i^\dna$.
Because all the possible states of the ring after scattering 
are orthogonal to each other, the outgoing channels are well defined and
the current is conserved. 

We calculate the eigenstates $\ket{\alpha_i^\sigma}$ 
using exact diagonalisation of the Hamitonian
$H_\rng$ in the suitable region of the many-particle Hilbert space, 
taking into account that the Hubbard Hamiltonian is invariant with respect 
to rotations in the spin space. The diagonalisation is therefore
performed in the constant $(n,S_z)$ space, where $S_z$ is the conserved
component of the total spin in the $z$ direction. The method
can be applied to Hamiltonians that don't conserve $S_z$ at the expense
of significantly more time consuming numerical calculations.

At zero temperature, the electron scatters on the ground state
of the n-particle state in the ring, $\ket{\alpha_0^\upa}$.
During the scattering the electron can loose energy
by exciting the bound electrons into one of the excited  $\ket{\alpha_i^\sigma}$
states. The probability of such transitions is a rapidly decreasing
function of the energy loss, therefore only a small number of
the scattering channels (states $\ket{\alpha_i^\sigma}$) has to
be considered. This observation is essential for numerical performance
of the method: we can efficiently calculate the states from the
bottom of the spectrum of the matrix representations of $H$ in suitable
$(n,S_z)$ subspaces using the iterative Lanczos technique.
We have used the implicitly restarted Lanczos method, as
implemented in ARPACK package \cite{arpack}. The eigenvalues
and eigenvectors were computed to machine precision.

By taking into account only the allowed states, the wave-function that
describes the scattering of one electron on the AB ring is given by
\begin{equation}
\begin{split}
\ket{\Psi} = &
\sum_{i=1}^{\infty} \sum_{j,\sigma} d^\lll_{i,j,\sigma} \cpm{i}{\sigma} \ket{\alpha_j^\sigma}
+\sum_{i=1}^{\infty} \sum_{j,\sigma} d^\rr_{i,j,\sigma} \cpn{i}{\sigma} \ket{\alpha_j^\sigma} \\
& + \sum_j e_j \ket{\beta_j},
\end{split}
\label{psi}
\end{equation}
where $d^L_{i,j,\sigma}$, $d^R_{i,j,\sigma}$ and $e_k$ are the coefficients
to be determined.

\subsection{Reduction to a sparse system of linear equations}

We consider a steady-state scattering described by the
the Schr\"odinger equation 
\begin{equation}\label{sch}
H\ket{\Psi}=E\ket{\Psi}
\end{equation}
with $H=H_\rng+H_{\text{lead}}+H_c$.  Strictly speaking, this equation
cannot be solved in the restricted space, spanned on the
$\ket{\alpha_i^\sigma}$ and $\ket{\beta_i}$ states, because applying
the Hamiltonian to the wave-function ansatz takes us out of this space
by generating terms where more than one electron can be found outside
the scattering region. Omission of these terms represents the main
approximation used in our method. This approximation leads to an error
which is not significant for suitably chosen model parameters (see
below).

Operating on the equation \eqref{sch} from the left 
with $\bra{\beta_l}$ we obtain
\begin{equation}
\label{brezprune}
-t_0 \sum_{j,\sigma} b^\lll_{l,j,\sigma} d^\lll_{1,j,\sigma} 
-t_0 \sum_{j,\sigma} b^\rr_{l,j,\sigma} d^\rr_{1,j,\sigma}
+ \sum_{k} h_{l,k} e_k = E e_l,
\end{equation}
where  $b$'s denote scalar products
\begin{equation}
\begin{split}
b^\lll_{l,j,\sigma}&=\bra{\beta_l} \cp{1}{\sigma}
\ket{\alpha_j^\sigma}, \\ 
b^\rr_{l,j,\sigma}&=\bra{\beta_l}
\cp{4}{\sigma} \ket{\alpha_j^\sigma},
\end{split}
\end{equation}
while $h_{l,k}=\braket{\beta_l|H_\rng}{\beta_k}$
are the matrix elements of Hamiltonian $H_\rng$ in the $n+1$ electron
subspace. 

By operating with $\bra{\alpha_j^\sigma} \cm{1}{\sigma}$ from the left
we get
\begin{equation}
\label{pred}
-t_{\text{lead}} d^\lll_{2,j,\sigma} 
-t_0 \sum_k (b^\lll_{k,j,\sigma})^\ast e_k + 
\epsilon_j^\sigma d^\lll_{1,j,\sigma} = E d^\lll_{1,j,\sigma}.
\end{equation}

\subsection{Pruning the leads}

In an open outgoing channel $(j,\sigma)$ a plane wave can propagate, 
so that $d^\lll_{2,j,\sigma}=\exp(i k_{j,\sigma}) d^\lll_{1,j,\sigma}$.
By   energy conservation the wave number $k_{j,\sigma}$ is obtained 
from 
\begin{equation}
\epsilon_0-2t_{\text{lead}} \cos(K)=
\epsilon^\sigma_j-2t_{\text{lead}} \cos(k_{j,\sigma}).
\end{equation}
The energy $\epsilon_0$ is the initial energy
of the $n$-electron bound state on the ring,
$K$ is the wave number of the incoming electron,
and $\epsilon^\sigma_j$ is the final energy of the bound electrons.
Equation~(\ref{pred}) can thus be written as 
\begin{equation}
d^\lll_{1,j,\sigma}=
\frac{-t_0 \sum_k (b^\lll_{k,j,\sigma})^\ast e_k}{  
E-\epsilon_j^\sigma+t_{\text{lead}} \exp(i k_{j,\sigma}) }.
\label{poprune1}
\end{equation}
Similar equation can be obtained for exponentially decaying (closed) 
outgoing channels that we also take into account (up to some cut-off
energy, above which the inclusion of further closed channels does
not change the results).
These are defined through the relation 
$d^\lll_{2,j,\sigma}=\exp(-\kappa_{j,\sigma}) d^\lll_{1,j,\sigma}$
and
\begin{equation}
\epsilon_0-2t_{\text{lead}}\cos(K)=
\epsilon^\sigma_j-2t_{\text{lead}} \cosh(k_{j,\sigma}).
\end{equation}
Equation (\ref{pred}) can in this case be written as
\begin{equation}
d^\lll_{1,j,\sigma}=
\frac{-t_0 \sum_k (b^\lll_{k,j,\sigma})^\ast e_k}{
E-\epsilon_j^\sigma+t_{\text{lead}} \exp(-\kappa_{j,\sigma}) }.
\label{poprune2}
\end{equation}
In the incoming channel we have both the incoming and outgoing waves,
$d^\lll_{m,0,\upa}=\exp(-i K m)+r \exp(i K m)$. We obtain
$d^\lll_{2,0,\upa}=\exp(i K) d^\lll_{1,0,\upa}+\exp(-2 i K)-1$.
The equation for the incoming channel thus contains 
an additional inhomogeneous term $\exp(-2 i K)-1$ and
Equation \eqref{pred} for the incoming channel is
\begin{equation}
d^\lll_{1,0,\upa}=
\frac{-t_0 \sum_k (b^\lll_{k,0,\upa})^\ast e_k 
-t_{\text{lead}}[\exp(-2iK)-1]
}{  
E-\epsilon_0+t_{\text{lead}} \exp(iK) }.
\label{poprune3}
\end{equation}

Using Eqs.~\eqref{poprune1}, \eqref{poprune2}, \eqref{poprune3},
and similar equations for the right lead, both leads can be removed
(pruned) from the problem \cite{bonca1}. 

Equations \eqref{brezprune}, \eqref{poprune1} and equivalent
equations for other outgoing channels 
form a system of linear equations for unknowns $d^\lll_{1,j,\sigma}$,
$d^\rr_{1,j,\sigma}$ and $e_j$.
This sparse system is solved for different energies of the incoming electron
using the SuperLU library \cite{superlu}.

The partial transmittivity through  channel $(j,\sigma)$ is given by
\begin{equation}
T_{j,\sigma}(E)=\frac{\sin(k_{j,\sigma})}{\sin(K)} |d^\rr_{1,j,\sigma}|^2.
\end{equation}
Since the method is based on exact solution of many-electron problem, 
we can compute transmission at arbitrarily large values of $U$.

\subsection{Extended interaction region}

Results can be improved by extending the interaction region which is
solved exactly by the Lanczos method by adding additional sites
from the leads. This procedure takes into account the decaying tails
of bound electron wavefunctions in the leads at the expense of
increasing the computational Hilbert space.
The error due to the omission of the terms corresponding 
to a second electron jumping out of the original interaction region 
[see discussion following
Eq.~\eqref{sch}] is exponentially reduced with the inclusion
of each additional site from the leads.

These improvements
mainly lead to energy shifts of the resonance peaks while the general
characteristics of the spectra remain unchanged. In principle, the
region could be extended until the desired convergence is achieved.
In our calculations the interacting region consisted of the ring and
one additional site from each lead, see Fig.~\ref{fig:ring}.
In fact, we had to take into account the site on the left 
electrode in order to ensure that the incident electron spreads 
into two partial waves that propagate through both arms of the ring. 
The additional site on the right electrode is required so that the partial
waves can interfere, which leads to the Aharonov-Bohm effect.
The inclusion of these two sites was therefore essential in
our studies of the dephasing mechanisms.

In cases where the ground state of the interaction region was degenerate, 
we averaged the transmittivity spectra over all the degenerate states.
The variational space taken into account in our calculation was equivalent 
to a Hubbard model on 8 sites with no translational symmetry. 

The largest problem that could be effectively solved has
3 bound electrons with spin up and 4 electrons with spin down.
In this case there are $\sim$8000 states $\ket{\alpha_i^\sigma}$
and $\sim$5000 $\ket{\beta_i}$ states. We kept 200 lowest
lying $\ket{\alpha_i^\sigma}$ states to define our scattering
channels (diagonalization took 3 minutes of a modest personal
computer).
Solving the sparse system of complex linear equations 
for a range of the incoming electron energies
took 200 minutes for 237 data points (or about
a minute per data point on the average).
This step is the most computationally demanding part of 
our technique. This is the main reason why we limited our calculations
to 8 site models, even though the Lanczos method easily
handles much larger lattices. 
\section{Results - one bound electron}

We now investigate the effect of interactions on an electron as it
tunnels through the ring. The incoming electron has spin up
and there is one bound electron with spin down inside the ring.
The on-site energies are $\epsilon=-4.5t_{\text{lead}}$, 
the overlap integrals are $t=\sqrt{3}t_{\text{lead}}$, 
and we set $t_{\text{lead}}=1$.  

\begin{figure}[htbp]
\includegraphics[angle=-90,clip,totalheight=6.6cm]{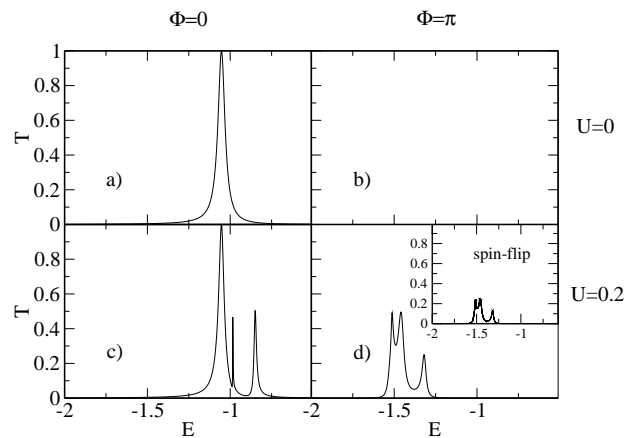}
\caption{Transmission probability as a function of the incident
electron energy for one electron with spin down bound
on the ring. The incoming electron had spin up. The coupling to
the lead is  $t_0=0.4$. In all cases transmission is purely elastic.
\label{fig:1}}
\end{figure}

First we consider the non-interacting case. In the absence of the 
magnetic field the transmission reaches unity at the resonance, Fig.~\ref{fig:1}a.
The electron is fully reflected at any incident energy when the magnetic
flux is $\Phi=\pi$, Fig.~\ref{fig:1}b. This is the usual
Aharonov-Bohm effect.

We now turn on the interaction. At $\Phi=0$ we still see  a unity peak 
at the energy of the single-electron resonance, followed by  smaller satellite
peaks caused by the  interaction, Fig.~\ref{fig:1}c. At $\Phi=\pi$,
when in the absence of $U$ the  electron is fully reflected,
we obtain very high 
transmission probability despite relatively  small $U=0.2$,
Fig.~\ref{fig:1}d.
In the largest peak the transmission approaches the value  $T=1/2$.
Since the incoming electron and the bound electron are not entangled,
their total spin is not well defined, therefore 
the total wavefunction is a superposition of a singlet and a triplet
state with equal amplitudes: $\ket{\uparrow \downarrow}=1/\sqrt{2}
\left (\ket{S=1,S_z=0} +  \ket{S=0,S_z=0}\right)$.
The triplet scattering has zero transmission probability at $\Phi=\pi$
since in the Hubbard model only singlet electrons interact.
The singlet scattering, however,  reaches the unitary limit at the main 
resonance peak. Averaging over both contributions, we indeed get $T=1/2$.

The spin-flip scattering part
of the transmission probability is shown in
the inset in Fig.~\ref{fig:1}d.  The spin-flip and normal scattering 
contribute equally to the total transmission probability. 
Both are purely elastic with respect to energy changes. 

\subsection{Lead decoupling at $\Phi=\pi$ and scattering mechanisms at non-zero $U$}

To gain more insight in the mechanism of non-zero electron tunneling
probability, we present a simple physical picture of electron
tunneling for the case of $\Phi=\pi$. We first transform the A-B ring
Hamiltonian \eqref{hring} from the basis of localized states to a
basis in k-space. For more generality, we can now assume that the ring
consists of an arbitrary even number $m$ of sites, which we now number
from 0 to $m-1$, so that the site 0 is coupled to the left electrode,
while the site $m/2$ is coupled to the right electrode.

\begin{figure}[htbp]
\includegraphics[height=5.5cm]{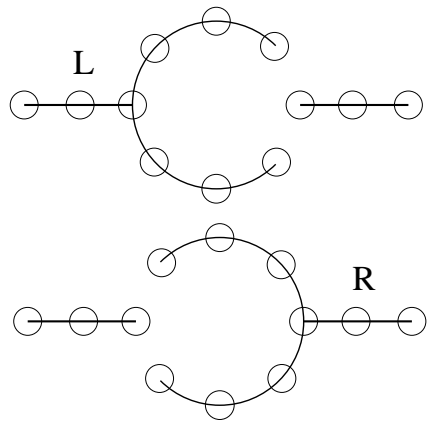}
\caption{At $\Phi=\pi$, one can rotate each pair of the degenerate eigenstates 
in such a way that one of them couples only to the left electrode and the other 
to the right electrode. In absence of interactions both electrodes are 
effectively decoupled. 
\label{decoup}}
\end{figure}

The non-interacting part of the Hamiltonian \eqref{hring} is diagonal in
the plane-wave basis,
\begin{equation}\label{travel}
d_{n,\sigma}=\frac{1}{\sqrt{m}} \sum_{j=1}^m e^{-ijk_n} c_{j,\sigma}
\end{equation}
with wave numbers $k_n$ given by the periodic boundary condition $\exp(ik_nm)=1$,
or $k_n=2\pi n/m$, where $n=0,\pm 1,\ldots,\pm m/2-1,m/2$.
The corresponding eigenvalues are
\begin{equation}\label{ener}
E_n=\epsilon-2t \cos(k_n-\Phi/m)
\end{equation}
When $\Phi=\pi$, all the non-interacting eigenstates are twofold
degenerate since $\cos(k_n-\fr{\pi}{m})=\cos(k_{1-n}-\fr{\pi}{m})$.
The complete orthonormal set of states is therefore composed of
$m/2$ pairs of states with wave-numbers $k_n$ and $k_{1-n}$ for $n$
ranging from $1$ to $m/2$.
For each pair we can form two linear combinations of states:
\begin{equation}\label{dlr}
\begin{split}
a_{L,n,\sigma}&=\fr{1}{\sqrt{2}}\left(d_{n,\sigma}+d_{1-n,\sigma}\right)\\
&=\fr{1}{\sqrt{2m}}\sum_j 
\left(e^{ik_nj}+e^{-ik_n j}e^{\fr{i2\pi j}{m}}\right)c_{j,\sigma}, \\
a_{R,n,\sigma}&=\fr{1}{\sqrt{2}}\left(d_{n,\sigma}-d_{1-n,\sigma}\right)\\
&=\fr{1}{\sqrt{2m}}\sum_j 
\left(e^{ik_nj}-e^{-ik_n j}e^{\fr{i2\pi j}{m}}\right)c_{j,\sigma}.
\end{split}
\end{equation}
It is easy to see that the coefficient of $c_{m/2,\sigma}$ in the expression
for $a_{L,n,\sigma}$ is zero and likewise for the coefficient of $c_{0,\sigma}$ 
in the expression
for $a_{R,n,\sigma}$. This means that the eigenstate denoted
by $L$ is coupled only to the left electrode, while eigenstate $R$ is
coupled only to the right electrode, see Fig.~\ref{decoup}. 
In the non-interacting case the
incoming electron can only tunnel from the left electrode to an $L$ state.
This state is decoupled from the right electrode and since there is no
term in the Hamiltonian, that would allow transitions from $L$ to $R$ state,
the electron is fully reflected. 

There are therefore two equivalent physical descriptions of zero
transmittivity of an AB ring. One can either consider it as a destructive
interference of partial electron waves that travel 
in the upper and lower arm of the ring, or as an effective decoupling of both 
electrodes due to a topological phase shift.

We will now write the interacting part of the Hamiltonian in the new
basis and search for processes that are responsible for the nonzero
transmission. From Eqs.~\eqref{travel} and \eqref{dlr} we can
express $c_{j,\sigma}$ as
\begin{equation}
c_{j,\sigma}=\fr{2e^{i\left(\fr{\pi j}{m}\right)} }{\sqrt{2m}} 
\sum_{n=1}^{m/2} \left( \cos(\tilde{k}_nj) a_{L,n,\sigma}
+ \sin(\tilde{k}_nj) a_{R,n,\sigma} \right).
\end{equation}
where we have introduced  shifted wave numbers 
\begin{equation}
\tilde{k}_n=\frac{2\pi}{m}\left(n-\frac{1}{2}\right).
\end{equation}

The particle number operator can then be expressed as
\begin{equation}
\begin{split}
c_{j,\sigma}^{\dag}c_{j,\sigma}=\fr{2}{m}\sum_{p,q}
&\cos(\tk_p j)\cos(\tk_q j)a_{L,p,\sigma}^{\dag}a_{L,q,\sigma} \\
+&\sin(\tk_p j)\sin(\tk_q j)a_{R,p,\sigma}^{\dag}a_{R,q,\sigma} \\
+&\cos(\tk_p j)\sin(\tk_q j)a_{L,p}^{\dag}a_{R,q,\sigma} \\
+&\sin(\tk_p j)\cos(\tk_q j)a_{R,p}^{\dag}a_{R,q,\sigma}.
\end{split}
\end{equation}
We now see that the Hubbard interaction term
$\sum_j \cp{j}{\upa} \cc{j}{\upa} \cp{j}{\dna} \cc{j}{\dna}$
is a sum over $j$ of products of four trigonometric
functions. Each one of these products can be written as a sum 
of trigonometric functions by using trigonometric reduction 
formulas such as, for example,
\begin{equation}
\begin{split}
8\sin(a)&\sin(b)\sin(c)\sin(d)=\\
&-\cos(a-b-c-d)+\cos(a+b-c-d) \\
&+\cos(a-b+c-d)-\cos(a+b+c-d) \\
&+\cos(a-b-c+d)-\cos(a+b-c+d) \\
&-\cos(a-b+c+d)+\cos(a+b+c+d).
\end{split}
\label{trigo}
\end{equation}
We note that reduction formulas for an even number of sine and cosine
functions consist of a sum of cosine functions, while the reduction
formulas for an odd number of sine and cosine functions consist of a
sum of sine functions. This fact is important to
understand the selection rules that lead to dephasing.

The arguments of functions on the right hand sides of the reduction
formulas are sums of the form
$\frac{2\pi j}{m}\left(p-\fr{1}{2}
\pm\left(q-\fr{1}{2}\right)
\pm\left(r-\fr{1}{2}\right)
\pm\left(s-\fr{1}{2}\right) \right)$,
i.e. of form $\frac{2\pi j t}{m}$,
where $t$ is an integer.
When the summation over site index $j$ is performed
most of the terms will drop, since
\begin{equation}
\begin{split}
&\frac{1}{m}\sum_{j=1}^m \cos\left( \frac{2\pi t}{m} j \right)=\delta_{t,0}, \\
&\frac{1}{m}\sum_{j=1}^m \sin\left( \frac{2\pi t}{m} j \right)=0.
\end{split}
\end{equation}

All interaction terms with a coefficient that after trigonometric
reduction involves a sine function will therefore vanish. Such
vanishing terms come from products of an odd number of trigonometric
functions of each kind, therefore they are of the form such as 
\begin{equation}
a_{R,p,\upa}^{\dag}a_{L,q,\upa} 
a_{L,r,\dna}^{\dag} a_{L,s,\dna}.
\end{equation}
Such terms would allow (for $p=q$, $r=s$) transitions
of the tunneling electron from state $L$ to $R$ without changing the
bound electron state (i.e. without leaving any imprint on the environment).
Such transitions would clearly be in contradiction with our understanding	
of the dephasing in the A-B rings. 

\begin{figure}[htbp]
\includegraphics[clip,totalheight=3cm]{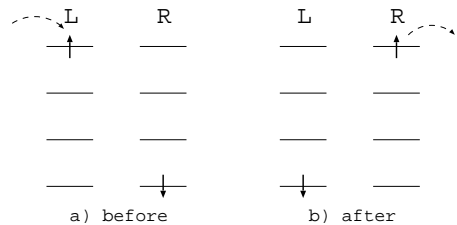}
\caption{Transmission due to elastic symmetry-changing scattering. 
\label{dg1}}
\end{figure}

\begin{figure}[htbp]
\includegraphics[clip,totalheight=3cm]{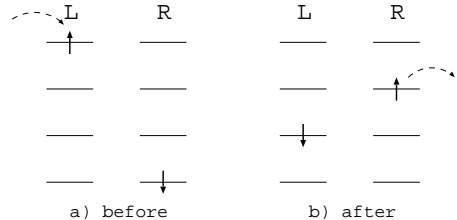}
\caption{Transmission due to inelastic symmetry-changing scattering. 
\label{dg2}}
\end{figure}

The terms with four sine and  with four cosine functions are of
little importance for our purposes. They describe interlevel repulsion 
and interlevel transitions without changes of the L/R character of the electron
states and therefore do not lead to a finite transmission. 
We will focus instead on terms with two sine and two cosine functions.
They are of three kinds. The first one consists of terms of the
form 
\begin{equation}
a_{R,p,\upa}^{\dag}a_{L,q,\upa}
a_{L,r,\dna}^{\dag} a_{R,s,\dna}.
\end{equation}
These terms describe what we call {\sl symmetry-changing transitions}: the tunneling
electron (with spin up) in the $L$ state jumps to a $R$ state, while a bound
electron undergoes a transition from $R$ to $L$ state. 
Such a transition can either be elastic (with respect to the energy of the
tunneling electron) if $p=q$, $r=s$ (Fig.~\ref{dg1}), or inelastic (Fig.~\ref{dg2}). 

\begin{figure}[htbp]
\includegraphics[clip,totalheight=3cm]{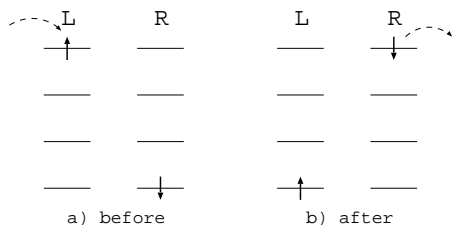}
\caption{Transmission due to elastic spin-flip scattering. 
\label{dg3}}
\end{figure}

\begin{figure}[htbp]
\includegraphics[clip,totalheight=3cm]{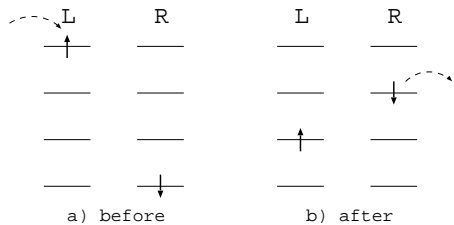}
\caption{Transmission due to inelastic spin-flip scattering. 
\label{dg4}}
\end{figure}

The second kind of terms is of the form
\begin{equation}
a_{L,p,\upa}^{\dag}a_{L,q,\upa} 
a_{R,r,\dna}^{\dag} a_{R,s,\dna}.
\end{equation}
These terms correspond to {\sl spin-flip transitions}: 
the tunneling electron with spin-up in the $L,q$ state makes
a transition to a lower laying state $L,p$ state, while a bound electron 
undergoes a transition from the $R,s$ to the
$R,r$ state, Fig.~\ref{dg3}. Transition can again be either elastic 
(with respect to energy) for $p=q$, $r=s$ (Fig.~\ref{dg3}), 
or inelastic (Fig.~\ref{dg4}). 

\begin{figure}[htbp]
\includegraphics[clip,totalheight=3cm]{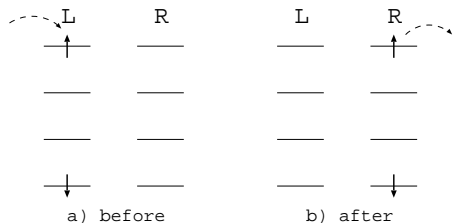}
\caption{Transmission due to elastic symmetry-changing scattering (of the second
kind). 
\label{dg5}}
\end{figure}

\begin{figure}[htbp]
\includegraphics[clip,totalheight=3cm]{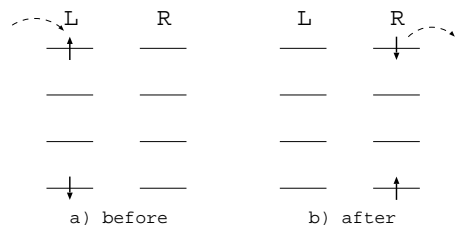}
\caption{Transmission due to elastic spin-flip scattering (of the second
kind). 
\label{dg6}}
\end{figure}

Finally, terms of the form 
\begin{equation}
a_{R,p,\upa}^{\dag}a_{L,q,\upa}
a_{R,r,\dna}^{\dag} a_{L,s,\dna}
\end{equation}
can correspond either to symmetry-changing (Fig.~\ref{dg5}) or to spin-flip 
transitions (Fig.~\ref{dg6}), depending on the $p,q,r,s$ quantum numbers. 

\subsection{Scattering of a wave packet}

To illustrate more in detail our results presented in the previous
subsection, we present here numerically exact calculation of
scattering of an incoming electron (described as a wave packet with a
given finite energy width) on the electron with the opposite spin,
bound on the Aharonov-Bohm ring. Since we are dealing with a simple
case of only two electrons, this problem can be solved numerically
exactly by direct integration of the two-body Schr\"odinger equation
$i\hbar d\ket{\psi}/dt=H\ket{\psi}$.  We take into consideration a
sufficiently high number of chain sites, so that the positional spread
of the wave packet is smaller than the length of the left and
the right lead. We chose $N=200$ sites, where the 6-site Aharonov-Bohm
ring occupies positions ranging from 101 to 106.

We construct the wavefunction at the initial time as
$\ket{\psi}=\psi^{\dag}_{\uparrow \mathrm{packet}} \phi^{\dag}_{\downarrow
\mathrm{bound}} \ket{0}$.  The operator $\phi^{\dag}_{\downarrow
\mathrm{bound}}$ creates an electron with spin down in the bound
eigenstate of the Aharonov-Bohm ring. We calculated this state using
direct diagonalisation. The operator $\psi^{\dag}_{\uparrow
\mathrm{packet}}$ is
\begin{equation}
\psi^{\dag}_{\uparrow\mathrm{packet}}=C \sum_k \exp\left[ - \frac{(k-k_0)^2}{2\sigma^2} \right]
\exp(-i k N_{\mathrm{center}}) c^\dag_{\uparrow k}
\end{equation}
where $c^\dag_{\uparrow k}=1/\sqrt{N}\sum_{j=1}^N \exp(i k j)
c^\dag_{\uparrow j}$
and $C$ is a normalization constant.
This operator creates an electron with spin up in a wave packet
centered at site $N_{\mathrm{center}}$ which has the average
wave number $k_0$ and a spread of $\sigma$ in the k-space. 
We choose $k_0=\pi/2$ to place the  wave packet 
in the middle of the energy band of the leads with  
the group velocity  $v=\partial E/
\partial k(k=k_0)=t=1$.  We set $\sigma=0.13$ and 
$N_\mathrm{center}=50$.

The equation of motion was then integrated using Bulirsch-Stoer
method, which gives highly accurate results for this type of
problems. The accuracy and stability can be conveniently
estimated by monitoring the deviation from the proper normalization 
of the wave-function. Using the Bulirsch-Stoer method the normalization
differs from 1 at the eighth decimal place after the scattering.

\begin{figure}[htbp]
\includegraphics[height=8cm,angle=-90,clip]{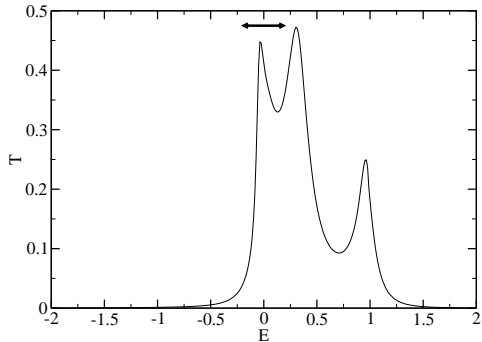}
\caption{Transmission probability as a function of the incident
electron energy for one electron with spin down bound on the ring.
\label{figU1}}
\end{figure}

We set the parameters to $\epsilon=-3.0$, $t=\sqrt{3}$, $t_0=0.6$,
$t_{\mathrm{lead}}=1$. For the non-interacting system the
transmittivity at $\Phi=\pi$ is 0 for all electron energies, while the
transmittivity of an interacting system with $U=1$ is shown in
Fig.~\ref{figU1}. The location and the spread of the energies of the
wavepacket are represented in the figure by a two-sided arrow.

\begin{figure}[htbp]
\includegraphics[height=8cm,angle=-90,clip]{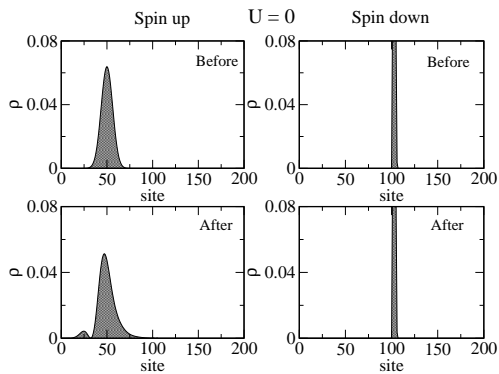}
\caption{Electron density before and after the scattering of the
wave-packet on an Aharonov-Bohm ring at $\Phi=\pi$: non-interacting case.
Note that the vertical scale is the same for both spin projections: the scale
was chosen so that the wave-packet is clearly visible. 
\label{slika1}}
\end{figure}

\begin{figure}[htbp]
\includegraphics[height=8cm,angle=-90,clip]{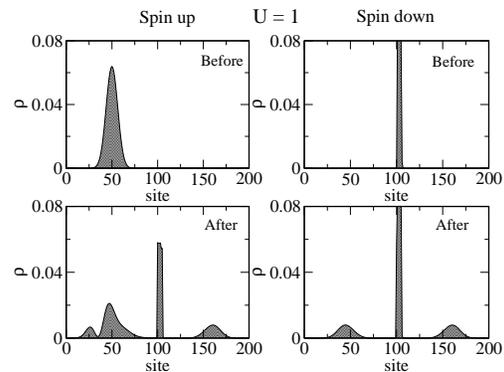}
\caption{Electron density before and after the scattering of the
wave-packet on Aharonov-Bohm ring at $\Phi=\pi$: interacting case.
\label{slika2}}
\end{figure}

\begin{figure}[htbp]
\includegraphics[height=8cm,angle=-90,clip]{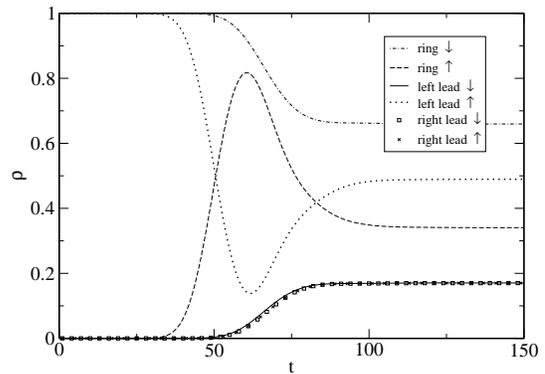}
\caption{
Time dependence (for interacting case) of the probability to find an electron 
with given spin projection either in the ring, in the left electrode or in
the right electrode. 
\label{cas}}
\end{figure}

The electron density before and after the scattering at $\Phi=\pi$ is
shown in Fig.~\ref{slika1} for the non-interacting case and in
Fig.~\ref{slika2} for the interacting case
\footnote{Animations of the scattering process are available 
upon request from the authors.
}.
For $U=0$, the wave-packet is perfectly reflected, as
expected. For $U=1$ the wave-packet is partially transmitted through
the scattering region. In fact, the expectation value to find an
electron in the second electrode, $P_R$, corresponds to the following
average:
\begin{equation}
P_R=\int \mathrm{d}k\, T(\epsilon(k))\vert\psi^\dagger_{\uparrow
\mathrm{packet}}(k,t=0)\vert^2,
\end{equation}
where the transmission $T$ is calculated using the method from section
II and is presented in Fig.~\ref{figU1}. This equation connects and thus
validates the two distinct methods. It is furthermore worth stressing,
that the probability of finding electrons with either orientation of
spin in the second electrode is equal, see Fig.~\ref{cas}. This can
be explained as follows: finite transmission is a direct consequence
of interaction $U$ which in the case of two electrons acts only on the
singlet part of the wavefunction. The triplet part does not feel $U$
due to the on-site nature of the interaction. Transmission therefore
occurs only through the singlet channel.

\subsection{Aharonov-Bohm oscillations}

Aharonov-Bohm effect is experimentally observed as magnetic flux
dependent oscillations of the electric current through a mesoscopic
ring structure \cite{webb}. From calculated $T(E)$ spectra we could
estimate the zero-bias conductance as $G=G_0 T(E_F)$, where
$G_0=2e^2/h$ is the conductance quantum and $E_F$ is the common Fermi
level of both leads. In our minimal model with a discrete number of
resonance states in the ring, the energy shifts of the peaks when the
flux is changed (see Eq.~\eqref{ener}) lead to pronounced conductance
variations not necessarily connected to the Aharonov-Bohm effect
itself. This is a direct consequence of using a small discrete number
of sites. It is therefore more revealing to observe the variations of
the {\sl integral} of transmittivity over the whole energy band,
$\int T(E) dE$. This
quantity is relatively insensitive to energy shifts of the peaks,
while it should clearly show A-B oscillations which affect the height
of all of the peaks.

\begin{figure}[htbp]
\includegraphics[clip,angle=-90,totalheight=6.5cm]{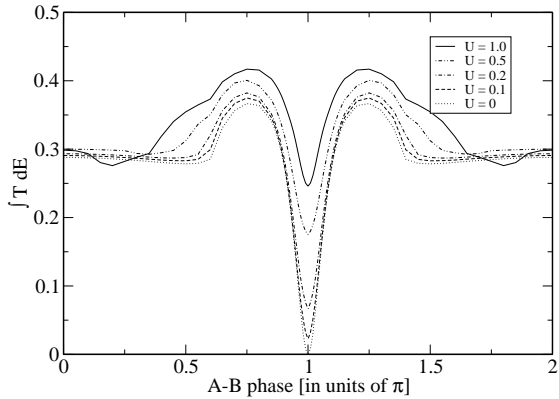}
\caption{
The flux dependence of integrated transmission probability for 
different interaction strengths. The coupling to the leads is
$t_0=0.6$, all other parameters are as before.
\label{fig:phi}
}
\end{figure}

In Fig.~\ref{fig:phi} we present this integral as a function of $\Phi$
for a number of increasing interaction strengths $U=0,0.1, \dots,1.0$.
The amplitude of A-B oscillations noticeably decreases as the
interaction grows stronger. Figure~\ref{fig:phi} also shows, that the
integral transmittivity is essentially interaction independent around
zero flux, $\Phi=0$. A similar insensitivity of the transmittivity
sum-rule has been discovered in the case of tunneling in the presence
of electron-phonon coupling \cite{bonca3,wingreen1,wingreen2}.
This insensitivity breaks down at larger U.

\section{Many bound electrons on the ring}

\begin{figure}
\includegraphics[angle=-90,clip,totalheight=6.6cm]{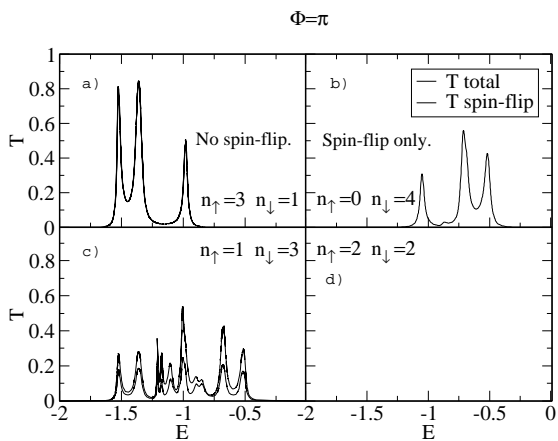}
\caption{Transmission probability as a function of the incident
electron energy for $n_\upa$ ($n_\dna$) electrons with spin up (down).
Parameters are $\epsilon=-4.5$, $t=\sqrt{3}$, $U=1.0$, $t_0=0.3$ and
$t_\mathrm{lead}=1$. \label{fig:2}}
\end{figure}

We now consider several interacting ($U=1$) bound
electrons on the ring. All presented cases 
are calculated at the flux value $\Phi=\pi$,
unless otherwise specified. Spin of the incoming electron is up.
We have limited the energy of the incident electron to a half of the
bandwidth, i.e. $E=[-2,0]$, in order to avoid ionisation.
Our main goal in this
section is to investigate the circumstances, under which a scattering
electron obtains a finite transmission probability at $\Phi=\pi$ when
scattering through the AB ring in the presence of many bound
electrons. We show that in most cases Coulomb interaction
leads to finite transmission. In our work we refer to processes which
cause finite transmission as {\it dephasing processes}. To avoid
confusion we point out once more that the total wavefunction describing a
many-body state of the scattering electron and bound electrons
preserves its full quantum coherence throughout the calculation.  Our
Hamiltonian does not contain coupling to external degrees of freedom
that would naturally lead to dephasing.

When the bound state on the ring consists of three electrons with spin
up and one electron with spin down (Fig.~\ref{fig:2}a) 
no spin-flip scattering is possible because such
processes turn out to be energetically impossible. The ground state is
however fourfold degenerate and the tunneling electron can get through
the ring at finite $U$ by changing the symmetry of the many-electron
state on the ring. Since the ground state is degenerate, this process
is purely elastic.

In the case of $n_\upa=0, n_\dna=4$ (Fig.~\ref{fig:2}b) 
the ground state is non-degenerate,
however the spin-flip processes are energetically allowed.  We
therefore obtain transmission probability only in spin-flipped
channels. Since in this case the ground state is not degenerate, the
transmission consists of purely inelastic processes.

In the case when the ground state is degenerate and the spin-flip
processes are allowed, we expect dephasing to occur both with or
without spin flip. Such is the case of $n_\upa=1, n_\dna=3$, Fig.~\ref{fig:2}c. The
transmittivity without spin-flip is purely elastic, while the
spin-flip processes are predominantly elastic, with small contribution
from inelastic channels.

Finally, for $n_\upa=2, n_\dna=2$ electrons are fully reflected from
the ring since there are no allowed scattering channels in the
appropriate energy interval, Fig.~\ref{fig:2}d.

\begin{figure}[htbp]
\includegraphics[clip,totalheight=11cm]{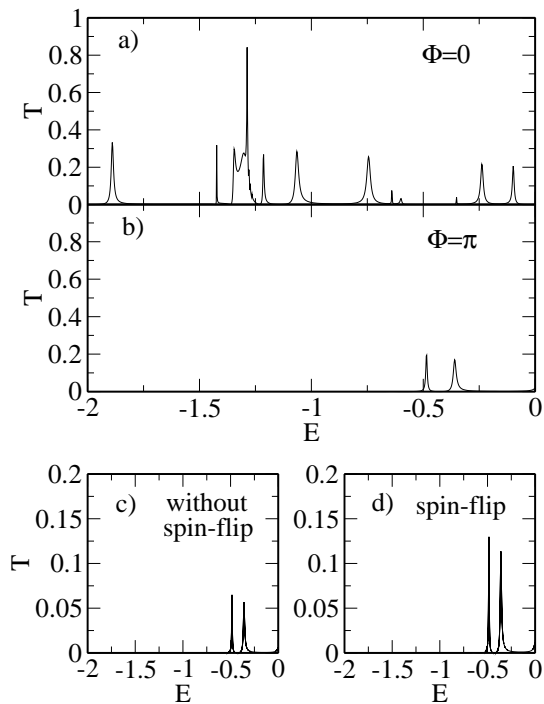}
\caption{Transmission probability as a function of incident
electron energy for  $n_\upa=2, n_\dna=2$,  $U=15$ and $\epsilon=-20$.
\label{fig:3}}
\end{figure}

We finally show the influence of large $U=15$ on the case of $n_\upa=2,
n_\dna=2$, where at $U=1$ 
transmission remained zero in the whole interval of incoming electron
energy due to widely spaced many-electron levels.
At large $U=15$ the energy difference between the
non-degenerate ground state and the first excited state decreases in
comparison with $U=1$ case, as the states become compressed in the
lower Hubbard band.  We changed the on-site energy to $\epsilon=-20$
in order to keep the electrons bound on the ring.  At $\Phi=0$ there
are several energies at which the electron can resonantly tunnel
through the ring, Fig.~\ref{fig:3}a.  At $\Phi=\pi$, the electron
can only tunnel inelastically.  The energy difference to the first
excited state in the $n$ electron Hubbard band is approximately
$1.4$. We find indeed that only the electrons that are more than $1.4$
above the bottom of the energy band can tunnel,
Fig.~\ref{fig:3}b. Such inelastic processes occur both without
(Fig.~\ref{fig:3}c) or with spin-flip (Fig.~\ref{fig:3}d).

\section{Conclusions}

Using a simple model and a new numerical method we have investigated
physics of single electron tunneling through the AB ring in the
presence of correlated bound electrons.  In particular we have focused
on the the role of electron-electron interactions on
dephasing. While the proposed method clearly has some limitations
(small interacting regions, inability to describe ionization
processes, neglect of many-body effects in the leads), it nevertheless
allows to treat the strong-interaction problem exactly and to identify
the two principal microscopic mechanisms which lead to the loss of
phase coherence in quantum interference experiments.  We showed that a
particle can tunnel through AB ring at $\Phi=\pi$ elastically by a)
changing the symmetry of the many-electron state which is possible in
the case of degeneracy or b) by flipping the spin.  Tunneling can also
occur in the inelastic channel by exciting the many-electron state on
the ring into an excited state with or without the spin-flip.
Depending on the number of bound electrons, their total spin,
degeneracy of the ground state and available energy of the incoming
electron, the total transmission can be composed of partial
transmissions caused by either one of the listed processes.

Using the novel method we have thus unraveled microscopic
mechanisms based on electron-electron interaction that in a mesoscopic
system contribute to a finite transmission through the AB ring in the
case of $\Phi=\pi$.  However, since our method is based on small
physical systems that can handle only a few lattice sites and
interacting electrons
we have no means at this stage to perform accurate calculation 
of the dephasing rate.

Even though all presented results are obtained on the basis of
zero-temperature calculations, the method can be generalised to finite
temperatures with some additional numerical effort.  On the other
hand, our results do not necessarily predict a finite dephasing rate
at zero temperature. Since we treat only a single electron in the
leads we are completely neglecting the effects of many-body
interactions spreading from the interacting region to the electrons in
the leads. This spread forms the basis for the Kondo effect. At
temperatures below the Kondo temperature $T_K$, our approach therefore
breaks down; in the Kondo regime the spins of the electrons from the
interacting region couple into singlets with the electrons from the
leads.  This process prevents spin-flip scattering, which in our
calculation represents one of the mechanisms for dephasing. Kondo
coupling may also lift the degeneracy of the many-electron states in
the interacting region and thus prevent transmission through the
elastic channel which  leads to dephasing at zero temperature
according to our findings. Other mechanisms leading to dephasing in
our approach might as well be modified in this low-temperature regime.
We  therefore conclude that despite the zero-temperature formalism
used in our method, our calculations are relevant only at temperatures
higher than the Kondo temperature $T_K$.

The method can be applied to study other many-body effects that are
expected to be important in nanoscopic structures due to strong
electron-electron and electron-phonon coupling. A more general
implementation of the presented method is under way.

\begin{acknowledgments}
We gratefully acknowledge S.A. Trugman, I. Martin, A. Ram\v sak,
T. Rejec and J. H.  Jefferson for fruitful discussions.  Authors also 
acknowledge the support of the Ministry of Education, Science and
Sport of Slovenia.
\end{acknowledgments}

\bibliography{ring}

\end{document}